\documentstyle[aps,12pt,epsf,amssymb]{revtex}
\begin{document}
\baselineskip=0.8cm
\newcommand{\ini}{\begin{equation}}
\newcommand{\fin}{\end{equation}}
\newcommand{\inir}{\begin{eqnarray}}
\newcommand{\finr}{\end{eqnarray}}
\newcommand{\inif}{\begin{figure}}
\newcommand{\finf}{\end{figure}}
\newcommand{\bc}{\begin{center}}
\newcommand{\ec}{\end{center}}
\def\ol{\overline}
\def\pa{\partial}
\def\ra{\rightarrow}
\def\ts{\times}
\def\df{\dotfill}
\def\bs{\backslash}
\def\dg{\dagger}

$~$

\hfill DSF-T-99/33

\vspace{1 cm}

\centerline{\LARGE{Testing quark mass matrices}}

\centerline{\LARGE{with right-handed mixings}}

\vspace{1 cm}

\centerline{{\large{D. Falcone}}$^1$ {\large{and F. Tramontano}}$^{1,2}$}

\vspace{1 cm}

\centerline{$^1$Dipartimento di Scienze Fisiche, Universit\`a di Napoli,}
\centerline{Mostra d'Oltremare, Pad. 19, I-80125, Napoli, Italy;}
\centerline{$^2$INFN, Sezione di Napoli, Napoli, Italy}

\centerline{ e-mail: falcone@na.infn.it}
\centerline{ e-mail: tramontano@na.infn.it}

\vspace{1 cm}

\begin{abstract}

\noindent
In the standard model, several forms of
quark mass matrices which correspond to the choice of weak bases lead to
the same left-handed mixings $V_L=V_{CKM}$, while the right-handed mixings
$V_R$ are not observable quantities. Instead, in a left-right
extension of the standard model, such forms are
ansatze and give different right-handed mixings which are now observable
quantities. We partially select the reliable forms of quark mass matrices
by means of constraints on right-handed mixings in some left-right models,
in particular on $V^R_{cb}$. Hermitian matrices are easily excluded. 

$~$

\noindent
PACS numbers: 12.15.Ff, 12.10.Dm

\end{abstract}

\newpage

\noindent
In the framework of the Standard Model (SM), based on the gauge group
$SU(3)_c \times SU(2)_L \times U(1)_Y$, the right-handed mixings are not
observable quantities, but they become observable in extensions of the SM
such as the Left-Right Model (LRM)
$SU(3)_c \times SU(2)_L \times SU(2)_R \times U(1)_{B-L}$ \cite{lr},
the Pati-Salam model
$SU(4)_{PS} \times SU(2)_L \times SU(2)_R$ \cite{ps}, and the grand
unified model $SO(10)$ \cite{10}.
Right-handed mixings are the most direct tool to test models of quark
mass matrices. Let us explain how this may happen.
In the LRM the quark mass and charged current terms are \cite{lansa}
\ini
\ol u _L M_u u_R+ \ol d _L M_d d_R+ g_L \ol u_L d_L W_L + g_R \ol u_R d_R W_R.
\fin 
Diagonalization of $M_u$, $M_d$ by means of the biunitary transformations
$$
U_u^{\dg} M_u V_u= D_u,~~U_d^{\dg} M_d V_d= D_d
$$
gives (renaming the quark fields)
\ini
\ol u _L D_u u_R+ \ol d _L D_d d_R+ g_L \ol u_L V_L d_L W_L +
g_R \ol u_R V_R d_R W_R,
\fin 
where
$$
V_L=U_u^{\dg} U_d=V_{CKM},~~V_R=V_u^{\dg} V_d
$$
are the left- and right-handed mixing matrices of quarks, and $D_u$, $D_d$
have non-negative matrix elements. In the SM the last term in Eqns.(1),(2)
is absent and it is possible to perform, without physical consequences,
that is without changing the observable quantities appearing in Eqn.(2),
the following unitary transformations on the quark fields:
\ini
u_L \ra {\cal U}u_L,~d_L \ra {\cal U}d_L,
\fin
\ini
u_R \ra {\cal V}_u u_R,~d_R \ra {\cal V}_d d_R.
\fin
In the LRM Eqn.(4) must be replaced by
\ini
u_R \ra {\cal V}u_R,~d_R \ra {\cal V}d_R,
\fin
that is also $u_R$ and $d_R$ must transform in the same way because of the
last term in Eqns.(1),(2). From the point of view of quark mass matrices,
the consequences of replacing Eqn.(4) with Eqn.(5), keeping Eqn.(3),
are the following.
In the SM we can use the freedom in ${\cal U}$ and ${\cal V}_u$
to choose $M_u=D_u$. Further we can use the freedom in ${\cal V}_d$ to
choose $M_d$ to be hermitian or to have three zeros
\cite{ma,fpr,hs,ft,kuo,fal}. In the LRM, the second freedom is not
there because both the diagonalizing matrices of $M_d$ are physical
observables.
This fact means that $bases$ $in~the$ $SM$ $become$ $ansatze$ $in~the$ $LRM$,
giving the same $V_L$ but different $V_R$.

The aim of this Letter is to begin a selection of quark mass matrices in
the LRM by using informations on right-handed mixings.
In fact, if without loss of generality one
sets $M_u=D_u$, then $V_L^{\dg} M_d V_R=D_d$, and thus
\ini
V_R^{\dg}=D_d^{-1} V_L^{\dg} M_d
\fin
permits to calculate the right-handed mixing matrix $V_R$ (values
of quark masses at the scale $M_Z$ and of the mixing $V_L$
are extracted from refs.\cite{fk} and \cite{rpp}).
It is well-known that if $M_d$ is hermitian or symmetric then $|V_R|=|V_L|$.
These conditions correspond to manifest and pseudomanifest left-right symmetry,
respectively \cite{lansa}.
In the general case, however, $V_R$ is not related to $V_L$ \cite{lansa}.
Notice that different quark mass and mixing matrices,
which correspond to bases in the SM, are connected by a
suitable unitary $U_R$, because
\ini
V_L^{\dg} M_d V_R=V_L^{\dg} M_d U_R U_R^{\dg} V_R=V_L^{\dg} M'_d V'_R,
\fin
where $M'_d=M_d U_R$ and $V'_R=U_R^{\dg} V_R$.
Therefore, they give different right-handed mixings in the LRM.
For example, let us consider the simple case of the first
two generations with real mass matrices (in the LRM mass matrices are complex
in general, even for only two generations).
The left-handed mixings are given by
$$
V_L \simeq \left( \begin{array}{cc}
                  1 & \lambda \\
                 -\lambda & 1
                  \end{array} \right),~ \lambda=0.22.
$$
In the SM, using a right-handed rotation ${\cal V}_d$ it is possible to put one
zero in $M_d$ in any position. In the LRM such four forms give different $V_R$.
We can get all forms from just one, for example from \cite{fal}
$$
M_d \simeq \left( \begin{array}{cc}
                  0 & \sqrt{m_d m_s} \\
                 \sqrt{m_d m_s} & m_s
                  \end{array} \right) \Rightarrow 
V_R \simeq \left( \begin{array}{cc}
                  -1 & \lambda \\
                 \lambda & 1
                  \end{array} \right),
$$
by using in Eqn.(7) $U_R$ like
$$
\left( \begin{array}{cc}
        0 & 1 \\
        1 & 0        
      \end{array} \right),
\left( \begin{array}{cc}
                  c & s \\
                 -s & c
                  \end{array} \right).
$$
In fact
$$
M_d U_R =\left( \begin{array}{cc}
                  0 & \sqrt{m_d m_s} \\
                 \sqrt{m_d m_s} & m_s
                  \end{array} \right)
\left( \begin{array}{cc}
        0 & 1 \\
        1 & 0        
      \end{array} \right)=
\left( \begin{array}{cc}
                 \sqrt{m_d m_s} & 0 \\
                 m_s & \sqrt{m_d m_s}
                  \end{array} \right)=M'_d,
$$
$$
U_R^{\dg} V_R =\left( \begin{array}{cc}
        0 & 1 \\
        1 & 0        
      \end{array} \right)
\left( \begin{array}{cc}
                  -1 & \lambda \\
                 \lambda & 1
                  \end{array} \right)=
\left( \begin{array}{cc}
                 \lambda & 1 \\
                 -1 & \lambda
                  \end{array} \right)=V'_R.
$$
The mixing $V^R_{us}$ is small on the first basis and large on the second.
Moreover,
$$
\left( \begin{array}{cc}
                  0 & \sqrt{m_d m_s} \\
                 \sqrt{m_d m_s} & m_s
                  \end{array} \right)
\left( \begin{array}{cc}
                  c & s \\
                 -s & c
                  \end{array} \right)=
\left( \begin{array}{cc}
                  -\sqrt{m_d m_s}s & \sqrt{m_d m_s}c \\
                 \sqrt{m_d m_s}c-m_s s & \sqrt{m_d m_s}s+m_s c
                  \end{array} \right),
$$
and imposing the element 2-1 to vanish, we have
$$
c=\sqrt{\frac{m_s}{m_s+m_d}}\simeq 1,~~s=\sqrt{\frac{m_d}{m_s+m_d}}\simeq
\lambda,
$$
and the third basis
$$
\left( \begin{array}{cc}
                  0 & \sqrt{m_d m_s} \\
                 \sqrt{m_d m_s} & m_s
                  \end{array} \right)
\left( \begin{array}{cc}
                  \sqrt{\frac{m_s}{m_s+m_d}} & \sqrt{\frac{m_d}{m_s+m_d}} \\
                 -\sqrt{\frac{m_d}{m_s+m_d}} & \sqrt{\frac{m_s}{m_s+m_d}}
                  \end{array} \right) \simeq
\left( \begin{array}{cc}
                  -m_d & \sqrt{m_d m_s} \\
                 0 & m_s 
                  \end{array} \right),
$$
$$
\left( \begin{array}{cc}
        1 & -\lambda \\
        \lambda & 1        
      \end{array} \right)
\left( \begin{array}{cc}
                  -1 & \lambda \\
                 \lambda & 1
                  \end{array} \right) \simeq
\left( \begin{array}{cc}
                 -1 & 0 \\
                 0 & 1
                  \end{array} \right).
$$
And again we can get the fourth basis, from the third, through
$$
\left( \begin{array}{cc}
                  -m_d & \sqrt{m_d m_s} \\
                 0 & m_s
                  \end{array} \right)
\left( \begin{array}{cc}
        0 & 1 \\
        1 & 0        
      \end{array} \right)=
\left( \begin{array}{cc}
                 \sqrt{m_d m_s} & -m_d \\
                 m_s & 0
                  \end{array} \right),
$$
$$
\left( \begin{array}{cc}
        0 & 1 \\
        1 & 0        
      \end{array} \right)
\left( \begin{array}{cc}
                  -1 & 0 \\
                 0 & 1
                  \end{array} \right)=
\left( \begin{array}{cc}
                 0 & 1 \\
                 -1 & 0
                  \end{array} \right).
$$
Mixing is nearly zero on the third basis and nearly one on the fourth. 
In this way, also for more than two generations,
one can construct different bases in the SM, which are different
ansatze in the LRM, from just few of them.

Therefore, let us consider now three generations and label elements in $M_d$ as
$$
\left( \begin{array}{ccc}
        1 & 2 & 3 \\
        4 & 5 & 6 \\
        7 & 8 & 9
       \end{array} \right).
$$
There are several SM bases with three zeros in $M_d$ \cite{fal}.
For example zeros can
be put in positions 137 \cite{ft} and 236 \cite{hs}, 478 \cite{kuo},
124. The last form can be obtained from 137 by just relabeling
the family indices 2,3.
From bases 124, 137, 478 we can calculate fifty-four bases by means of the six
special rotations
$$
\left( \begin{array}{ccc}
        1 & 0 & 0 \\
        0 & 1 & 0 \\
        0 & 0 & 1
       \end{array} \right),
\left( \begin{array}{ccc}
        1 & 0 & 0 \\
        0 & 0 & 1 \\
        0 & 1 & 0
       \end{array} \right),
$$
$$
\left( \begin{array}{ccc}
        0 & 1 & 0 \\
        1 & 0 & 0 \\
        0 & 0 & 1
       \end{array} \right),
\left( \begin{array}{ccc}
        0 & 1 & 0 \\
        0 & 0 & 1 \\
        1 & 0 & 0
       \end{array} \right),
$$
$$
\left( \begin{array}{ccc}
        0 & 0 & 1 \\
        1 & 0 & 0 \\
        0 & 1 & 0
       \end{array} \right),
\left( \begin{array}{ccc}
        0 & 0 & 1 \\
        0 & 1 & 0 \\
        1 & 0 & 0
       \end{array} \right),
$$
which produce permutations of columns in $M_d$ and of rows in $V_R$,
and suitable unitary transformations of the type
$$
\left( \begin{array}{ccc}
        1 & 0 & 0 \\
        0 & e^{i \alpha}c & e^{i \beta}s \\
        0 & -e^{i \gamma}s & e^{i \delta}c
       \end{array} \right),~\alpha+\gamma=\beta+\delta.
$$
For each of the three starting
forms under examination we have calculated the matrix $M_d$ by the relation \cite{ft}
\ini
M_d M_d^{\dg}=V_L D_d^2 V_L^{\dg}
\fin
and $V_R$ by Eqn.(6). For $V_L$ we use the standard parametrization \cite{rpp}.
Moreover, to keep arbitrary representation of $V_L$ one must put three phases
(not just one SM observable) in $M_d$ \cite{fpr}. Their positions for our
starting bases are
356, 256, 236, respectively. Putting three phases and their position
is part of the ansatze in the LRM,
because for three generations of quarks there are seven observable phases,
one can be inserted in $V_L$ and six in $V_R$ \cite{herc}.
However, due to the position of the three zeros,
putting the three phases in another position, or more than three phases,
up to six, does not change the moduli of $V_R$.
From the three starting bases,
by means of the six rotations we get eighteen bases and with the help of the
unitary transformation we get other six bases (imposing one element in
the second column to vanish) which become thirty-six by using again the six
rotations, making a total of fifty-four.
These are all the SM bases with $M_u=D_u$ and $M_d$ containing three zeros,
out of eighty-four possibilities. We try to understand if some of the
fifty-four bases satisfy constraints coming from $B$ decay, $K_L-K_S$ mass
difference and $B- \ol{B}$ mixing, within the LRM.

\newpage

\begin{table}
\begin{center}
\begin{tabular}{c|cccccccc}
zeros & 124 & 236 & 146 & 256 & 149 & 259 & 127 & 128 \\ 
phases & 356 & 145 & 589 & 479 & 568 & 467 & 356 & 346 \\ 
$|V_{cb}^R|$ & 0.896 & 0.896 & 0.789 & 0.789 & 0.984 & 0.984 & 0.914 & 0.914 \\
\hline
\hline
zeros & 167 & 268 & 479 & 589 & 347 & 358 & 467 & 568 \\ 
phases & 589 & 479 & 235 & 134 & 256 & 146 & 235 & 134 \\ 
$|V_{cb}^R|$ & 0.785 & 0.785 & 0.999 & 0.999 & 0.875 & 0.875 & 0.871 & 0.871
\end{tabular}
$~$

Table 1. Position of zeros and phases in $M_d$ and corresponding prediction
for $|V_{cb}^R|$.
\end{center}
$~$

\noindent
In fact, a recent analysis \cite{riz} of right-handed currents in B decay within
the LRM suggests that
$|V_{cb}^R|$ is large and perhaps near unity. In such analysis
$M_{W_R} \gtrsim 720$ GeV \cite{d0} is supposed. Actually, this experimental
bound is obtained by manifest left-right symmetry.
From inclusive semileptonic decays of $B$ mesons ref.\cite{riz}
gives $|V_{cb}^R| \gtrsim 0.782$. Moreover, if, as suggested in ref.\cite{vol},
right-handed currents can help to solve the $B$ semileptonic branching fraction
and charm counting problems, then ref.\cite{riz} gives $|V_{cb}^R| \ge 0.908$
and $M_{W_R} \lesssim 1600$ GeV.
For our purposes we assume $|V_{cb}^R|>0.750$.
We have used this constraint
to select quark mass matrices, taking $\delta=75^{\circ}$ in $V_L$. This value
is well inside the experimentally favoured region \cite{rpp}.
A moderate variation of $\delta$, say from $60^{\circ}$ to $90^{\circ}$, has
not a relevant effect. 
Of course, the hermitian form of $M_d$ (and in general of both mass matrices)
is excluded because it yields $|V_{cb}^R|=|V_{cb}^L| \simeq \lambda^2$. 
The successful ansatze are in Table 1, where the position of phases can be
changed by a diagonal phase matrix $U_R$.
It should be noted that ansatze 149, 259, 167, 268, 347, 358 give the quite
particular exact result that some element in
$M_d^{\dg} M_d=V_R D_d^2 V_R^{\dg}$ is zero.
\end{table}

\noindent
As an example of a successful ansatz we report $|M_d|$ and $|V_R|$
for model 124:
$$
|M_d|=\left( \begin{array}{ccc}
        0 & 0 & 0.023 \\
        0 & 0.106 & 0.104 \\
        0.541 & 2.687 & 1.213
       \end{array} \right),~~
|V_R|=\left( \begin{array}{ccc}
        0.958 & 0.221 & 0.180 \\
        0.205 & 0.393 & 0.896 \\
        0.199 & 0.892 & 0.405
       \end{array} \right).
$$
The near equality of elements 5 and 6 in $|M_d|$ is due to the specific
value $\delta \simeq 75^{\circ}$ \cite{fal}. The matrix $|V_R|$ has an approximate
symmetric expression
$$
|V_R| \simeq \left( \begin{array}{ccc}
        1 & \lambda & \lambda  \\
        \lambda & 2\lambda & 1 \\
        \lambda & 1 & 2\lambda
       \end{array} \right).
$$
Further constraints on the form of $V_R$ come from $K_L - K_S$ mass difference
\cite{kk} and $B - \ol{B}$ mixing. Assuming that each row and column of $V_R$
contains only one large element and forbidding fine-tuning,
these constraints give \cite{riz}
$|V^R_{us}| \lesssim \lambda^2$, $|V^R_{ub}| \lesssim \lambda$,
$|V^R_{tb}| \lesssim \lambda$, when $|V^R_{ts}|$ is large, and
$|V^R_{ud}| \lesssim \lambda$, $|V^R_{ub}| \lesssim \lambda$,
$|V^R_{tb}| \lesssim \lambda^3$, when is $|V^R_{td}|$ large. 
Out of the sixteen models in Table 1, the 128 satisfies quite well the second
set of constraints:
$$
|M_d|=\left( \begin{array}{ccc}
        0 & 0 & 0.023 \\
        0.103 & 0.021 & 0.104 \\
        2.741 & 0 & 1.213
       \end{array} \right),~~
|V_R|=\left( \begin{array}{ccc}
        0.199 & 0.892 & 0.405 \\
        0.088 & 0.395 & 0.914 \\
        0.976 & 0.217 & .0003
       \end{array} \right).
$$
The approximate expression for $|V_R|$ is
$$
|V_R| \simeq \left( \begin{array}{ccc}
        \lambda & 1 & 2 \lambda  \\
        2 \lambda^2 & 2 \lambda & 1 \\
        1 & \lambda & \lambda^5
       \end{array} \right).
$$
Model 124 gives $|V^R_{us}| \simeq \lambda$ rather than
$|V^R_{us}| \lesssim \lambda^2$. Also models 479 and 589 are reliable,
with very small mixings.

To better explain the physical content of the foregoing calculation we present
some comments. We have considered fifty-four forms of quark mass matrices with
three zeros and three phases inside $M_d$ and a diagonal $M_u$.
As we said, $|V_R|$ does not change if 
we put more than three phases.
On the other hand the existence of
three zeros in $M_d$ is a strong restriction because in the LRM just one or two
zeros settle an ansatz. Nevertheless, we have found a few models that satisfy
the constraints from $K$ and $B$ physics. Other ansatze can be obtained
starting from a diagonal $M_d$.

We stress the simple result that, if $|V_{cb}^R|$ is large,
hermitian or symmetric
quark mass matrices \cite{rrr} are not reliable \cite{wz}. Non-symmetric
mass matrices have important applications in the leptonic sector, mainly in
connection with the large mixing of neutrinos \cite{ho}.

Using Eqns.(3),(5) it is possible to change the structure of both
$M_u$ and $M_d$. For example,
from model 128, multiplying to the right by a simple unitary transformation
in the 2-3 sector, it yields $M_d$ with one zero in position 1 and $M_u$ with
four zeros in positions 2347. Although such forms can be more interesting to
discover an underlying theory of fermion masses and mixings, they lead to
the same parameters of Eqn.(2), and we need other observable quantities to
make a selection of such models with non diagonal mass matrices.
Such new physical parameters exist in extensions of the LRM.
Actually, in the SM one can get $M_u$ diagonal and
$M_d$ with three zeros; in the LRM $M_u$ can be diagonalized but $M_d$ is
fixed. In the Pati-Salam model, due to the relation between quarks and leptons,
we cannot choose $M_u$ diagonal in the general case.
In $SO(10)$ also $u_L$ and $u_R$ transform in the same way and
then it is never possible
to choose $M_u$ diagonal. Non-symmetric mass matrices can be obtained by
using also the antisymmetric Higgs {\bf 120} in the Yukawa couplings with
fermions, or if one allows for effective non-renormalizable couplings of
the light generations \cite{adhrs}.

\newpage

In conclusion, for the first time, we have performed, within the LRM,
a systematic
study of quark mass matrices which have a general structure in the SM.
Constraints on right-handed mixings, coming from various experimental and
theoretical sources, permit to select three reliable forms, apart from phases.

$~$

The authors thank F. Buccella for comments on the manuscript and T. Rizzo
for communications.


\begin{thebibliography}{100}

\newpage

\bibitem{lr} R.N. Mohapatra and J.C. Pati, Phys. Rev. D {\bf 11}, 566 (1975);
{\bf 11}, 2558 (1975)

R.N. Mohapatra and G. Senjanovic, Phys. Rev. D {\bf 12}, 1502 (1975)

\bibitem{ps} J.C. Pati and A. Salam, Phys. Rev. D {\bf 10}, 275 (1974);
Phys. Rev. Lett. {\bf 31}, 661 (1973); Phys. Rev. D {\bf 8}, 1240 (1973)

\bibitem{10} H. Georgi, in $Particles~and~Fields$, ed. C.E. Carlson
(AIP, New York, 1975)

H. Fritzsch and P. Minkowski, Ann. Phys. {\bf 93}, 193 (1975)

\bibitem{lansa} P. Langacker and S.U. Sankar, Phys. Rev. D {\bf 40}, 1569
(1989) and references therein

\bibitem{ma} E. Ma, Phys. Rev. D {\bf 43}, R2761 (1991)

\bibitem{fpr} D. Falcone, O. Pisanti, and L. Rosa, Phys. Rev. D {\bf 57}, 195
(1998)

\bibitem{hs} R. Haussling and F. Scheck, Phys. Rev. D {\bf 57}, 6656 (1998)

\bibitem{ft} D. Falcone and F. Tramontano, Phys. Rev. D {\bf 59}, 017302 (1999)

\bibitem{kuo} T.K. Kuo, S.W. Mansour, and G.-H. Wu, Phys. Rev. D {\bf 60},
093004 (1999)

\bibitem{fal} D. Falcone, Mod. Phys. Lett. A {\bf 14}, 1989 (1999)

\bibitem{fk} H. Fusaoka and Y. Koide, Phys. Rev. D {\bf 57}, 3986 (1998)

\bibitem{rpp} Particle Data Group, C. Caso $et~al.$, Eur. Phys. J. C {\bf 3},
1 (1998), p. 103 

\bibitem{herc} P. Herczeg, Phys. Rev. D {\bf 28}, 200 (1983)

\bibitem{riz} T.G. Rizzo, Phys. Rev. D {\bf 58}, 055009 (1998); {\bf 58},
114014 (1998)

\bibitem{d0} D0 Collaboration, S. Abachi $et~al.$, Phys. Rev. Lett. {\bf 76},
3271 (1996); B. Abbott $et~al.$, International Europhysics Conference on
High Energy Physics, Jerusalem, Israel, 1997

CDF Collaboration, F. Abe $et~al.$, Phys. Rev. Lett. {\bf 74}, 2900 (1995)

\bibitem{vol} M.B. Voloshin, Mod. Phys. Lett. A {\bf 12}, 1823 (1997)

\bibitem{kk} G. Beall, M. Bander, and A. Soni, Phys. Rev. Lett. {\bf 48},
848 (1982)

G. Ecker and W. Grimus, Nucl. Phys. B {\bf 258}, 328 (1985)

\bibitem{rrr} Systematic studies on hermitian or symmetric mass matrices are
P. Ramond, R.G. Roberts, and G.G. Ross, Nucl. Phys. B {\bf 406}, 19 (1993);
M. Randhawa, V. Bhatnagar, P.S. Gill, and M. Gupta, Phys. Rev. D {\bf 60},
051301 (1999); T.K. Kuo, S.W. Mansour, and G.-H. Wu, Phys. Lett. B {\bf 467},
116 (1999); M. Baillargeon, F. Boudjema, C. Hamzaoui, and J. Lindig,
hep-ph/9808207

\bibitem{wz} Some early papers on non-hermitian and non-symmetric mass matrices
are F. Wilczek and A. Zee, Phys. Lett. B {\bf 70}, 418 (1977); G.C. Branco and
J.I. Silva-Marcos, Phys. Lett. B {\bf 331}, 390 (1994); T. Ito, Prog.
Theor. Phys. {\bf 96}, 1055 (1996)

\bibitem{ho} See for example K. Hagiwara and N. Okamura, Nucl. Phys. B
{\bf548}, 60 (1999); Z. Berezhiani and A. Rossi, JHEP {\bf 03}, 002 (1999);
G. Altarelli and F. Feruglio, Phys. Lett. B {\bf 451}, 388 (1999)

\bibitem{adhrs} G. Anderson, S. Dimopoulos, L.J. Hall, S. Raby, and G.D.
Starkman, Phys. Rev. D {\bf 49}, 3660 (1994)


\end{thebibliography}
\end{document}